\documentstyle[twoside,12pt,epsf]{article}
\normalsize
\newcommand{\bdi}{\begin{displaymath}}
\newcommand{\edi}{\end{displaymath}}
\newcommand{\bfi}{\begin{figure}}
\newcommand{\efi}{\end{figure}}

\newcommand{\beq}{\begin{equation}}
\newcommand{\eeq}{\end{equation}}
\newcommand{\beqa}{\begin{eqnarray}}
\newcommand{\eeqa}{\end{eqnarray}}

\def\longbar#1{\setbox1=\hbox{$#1$}
\setbox2=\vbox{\hrule width 0.8\wd1}
\raise0.5\ht1\hbox{${\lower\dp1\box2}\atop\box1$}}  

\begin{document}

\begin{titlepage}

\begin{flushright}
\today
\end{flushright}

\vspace{1cm}
\begin{center}
{\Large \bf Comment on covariant Stora--Zumino chain terms}\\[1cm]
C. Adam* \\
School of Mathematics, Trinity College, Dublin 2 \\

\vfill
{\bf Abstract} \\
\end{center}
In a recent paper, Ekstrand proposed a simple expression from which
covariant anomaly, covariant Schwinger term and higher covariant chain terms
may be computed. We comment on the relation of his result to the earlier
work of Tsutsui.

\vfill

$^*)${\footnotesize  
email address: adam@maths.tcd.ie, adam@pap.univie.ac.at} 

\end{titlepage}

\section{Introduction}

There are several methods for the computation of anomalies, anomalous
commutators (Schwinger terms) and higher terms of the Stora--Zumino chain, 
both for the consistent and covariant case. Among these are perturbative,
functional and algebraic methods \cite{Ad1}--\cite{COMANOM} (see
\cite{Bertl1} for a review).

One particularly wellknown and useful method for the computation of the
consistent anomalous chain terms is the method of descent equations a la
Stora and Zumino \cite{Sto1,Zu1}. 
In this framework, all consistent chain terms are computed
from the Chern--Simons form $\Omega_{2n-1} (A,F)$
with the help of some algebraic cohomology methods (here $A,F$ are
Lie-algebra-valued forms, $A=A^a_\mu T^a dx^\mu$, $F=dA+A^2$). It turns out 
that the anomalous chain terms (which are determined up to exact terms)
are just the expansion coefficients $\Omega^k_{2n-1-k}(v,A,F)$ of the
shifted Chern--Simons form
\beq
\Omega_{2n-1}(A+v,F)=\sum_{k=0}^{2n-1}\Omega^k_{2n-1-k}(v,A,F)
\eeq
in powers of $v$, where $v$ is the ghost field with ghost number 1.

For the covariant case, an analogous cohomological analysis was performed by
Tsutsui,
\cite{Tsu}. There, a procedure for computing all the covariant anomalous
chain terms was derived, and an expression in terms of a shifted
Chern--Simons form, analogous to (1), was given (see (11) below).

Recently, a simple expression for a shifted Chern--Simons form for the
covariant chain terms was proposed by Ekstrand \cite{Ek1} (see (6) below). 
Further, he proved
that his expression (i.e., the linear and quadratic expansion terms in $v$)
correctly reproduce the covariant anomaly and
Schwinger term.
In addition, he noticed that his expression deviates from the one given
by Tsutsui, \cite{Tsu}, already for the Schwinger term (i.e., at order
$v^2$).

It is the purpose of this paper to comment on that difference. We will show
that the algebraic construction of Tsutsui, \cite{Tsu}, is completely
equivalent to the proposed shift formula of Ekstrand, \cite{Ek1}, and
leads to the same covariant chain terms in all orders. The mentioned difference
is due to the fact that the explicit shift formula for the covariant chain
terms that was given in \cite{Tsu} is incorrect and does not reproduce the 
algebraic results of that paper.

\section{Covariant chain terms a la Ekstrand and Tsutsui}

As already mentioned, the (non-integrated) consistent anomaly, Schwinger
term and higher chain terms are given by the expansion in powers of the
ghost of the shifted Chern--Simons density
\bdi
\Omega_{2n-1}(A+v,F)=\Omega_{2n-1}(A+v,dA+A^2)
\edi
\beq
=\Omega_{2n-1}(A+v,(d+\delta)(A+v)+(A+v)^2)
\eeq
where $\delta$ is the BRS operator
\bdi
\delta A=-Dv \, , \quad \delta F=[F,v]\, , \quad \delta v=-v^2
\edi
\beq
[\delta ,d]=\delta^2 =0 
\eeq
\beq
D \equiv d+[A,\,\, ]
\eeq
(all commutators are graded w.r.t. form and ghost degree), and the Russian
formula \cite{Sto1,Zu1} ($\widehat A\equiv A+v$)
\beq
\widehat F(\widehat A)\equiv (d+\delta )(A+v) +(A+v)^2 =dA +A^2 \equiv F(A)
\eeq
has been used. It was observed in \cite{Ek1} that it is precisely the 
occurrence of the BRS operator $\delta$ in (2) that makes the resulting 
expressions for the anomaly and Schwinger term non-covariant. Therefore,
it was proposed in \cite{Ek1} that a similar expression for covariant chain
terms may be found by simply dropping $\delta$ in (2). Further, it was proven
that this proposition is correct, i.e., that the expression
($\Omega_{2n-1}(A,F)\equiv \bar \Omega_{2n-1}(A,dA)$)
\bdi
\Omega_{2n-1}(A+v,d(A+v)+(A+v)^2)=\bar \Omega_{2n-1}(A+v,d(A+v))
\edi
\beq
=\sum_{k=0}^{2n-1}\bar \Omega^k_{2n-1-k}(v,dv,A,F)
\eeq
has linear and quadratic (in $v$) contributions that agree with the covariant 
anomaly and Schwinger term as computed by other methods.

As we want to relate this result (6) to the cohomological computations of
\cite{Tsu}, we should briefly review the latter. In \cite{Tsu} the following
two even (w.r.t. their total grading) operators $m,l$ are introduced
\bdi
mA=v \, , \quad mF=0
\edi
\beq
mv=0 \, ,\quad mdv =-v^2
\eeq
\bdi
lA=0 \, ,\quad lF=-Dv
\edi
\beq
lv=0\, ,\quad ldv =-v^2 .
\eeq
Both $m$ and $l$ act algebraically on formal polynomials of $A$, $F$, $v$
and $dv$. The consistent chain terms may be recovered by the action of
$m$ alone. Indeed,
\beq
\Omega^k_{2n-1-k}(v,A,F)=\frac{1}{k!}m^k\Omega_{2n-1}(A,F).
\eeq
On the other hand, both $m$ and $l$ are needed for the computation of the
covariant chain terms. It was proved in \cite{Tsu} that the covariant
chain terms will indeed be covariant provided they are computed as
\beq
\bar \Omega^k_{2n-1-k}(v,dv,A,F)=\frac{1}{k!}(m-l)^k \Omega_{2n-1}(A,F).
\eeq
As a shift formula that should incorporate all these covariant chain terms,
the following (incorrect) expression was given in \cite{Tsu}
\beq
\Omega_{2n-1}(A+v,F+Dv).
\eeq
This expression does not reproduce (10) because $m$ and $l$ do not
commute. More precisely, the problem is that the algebraic restrictions
of $m$ and $l$ to derivations on $(A,F)$
(these restrictions commute with $v$ and $dv$)
do not commute which each other, because only these 
restrictions are relevant for the difference $m-l$. Even if $m$ and $l$ 
are changed w.r.t. their action on $(v,dv)$ so that they commute 
which each other as
derivations on $(A,F,v,dv)$, which may be achieved by choosing $mdv=ldv
=-2v^2$ instead of $-v^2$ like in (7), (8),  the shift formula (11) would
be wrong and the formula (12) below would be correct.

The correct shift
formula that takes this non-commutativity into account and reproduces (10), is
\beq
\exp (v\frac{\delta}{\delta A} +Dv \frac{\delta}{\delta F})\Omega_{2n-1}
(A,F)
\eeq
as may be checked easily
(see also \cite{COMANOM}, journal version).
Here the exponential is defined as its power series, and the derivative is 
understood in an algebraic sense, e.g., $v (\delta /\delta A) P(A)=
P(A+v)|_{v^1}$ (i.e., the component of $P(A+v)$ linear in $v$). This formula
is equal to expression (6) that was proposed in \cite{Ek1}, as may
be proved  easily. Indeed,
\bdi
e^{v\frac{\delta}{\delta A}+Dv\frac{\delta}{\delta F}}\Omega_{2n-1}(A,F)=
\edi
\bdi
e^{(Dv +v^2)\frac{\delta}{\delta F}}e^{v\frac{\delta}{\delta A}}
\Omega_{2n-1}(A,F)=
\edi
\bdi
\Omega_{2n-1}(A+v,F+Dv+v^2)=
\edi
\beq
\Omega_{2n-1}(A+v,d(A+v)+(A+v)^2)
\eeq
where we used the Baker--Campbell--Hausdorff formula in the first step.

\section{Summary}

We have shown that the ``shift formula'' (6) that was proposed in \cite{Ek1}
for the generation of covariant chain terms and the cohomological analysis
of \cite{Tsu} lead to identical results for all chain terms. The
discrepancy that was mentioned in \cite{Ek1} is simply due to an
incorrect expression for the shift formula in \cite{Tsu}, which is not
supported by the (correct) cohomological computation of the same paper.

Our finding further points towards the correctness of the shift formula
proposed in \cite{Ek1} for all covariant chain terms. It should be mentioned
at this point, however, that a physical interpretation of the higher
covariant chain terms has not yet been found.

\section*{Achnowledgement}
The author thanks the School of Mathematics at Trinity College, Dublin, 
where this work was performed, for their hospitality. 
Further thanks are due to C. Ekstrand for helpful comments.
The author is supported by a Forbairt Basic Research Grant.


\begin{thebibliography}{999999}

\bibitem{Ad1}
S. Adler, Phys. Rev. 177 (1969), 2426.
\bibitem{BJ1}
J. S. Bell and R. Jackiw, Nuovo Cim. A60 (1969), 47.
\bibitem{Ba1}
W. A. Bardeen, Phys. Rev. 184 (1969), 1848.
\bibitem{JJ1}
R. Jackiw and K. Johnson, Phys. Rev. 182 (1969), 1459.
\bibitem{Fuji1}
K. Fujikawa, Phys. Rev. Lett. 42 (1979) 1195.
\bibitem{Fuji2}
K. Fujikawa, Phys. Rev. D21 (1980) 2848. 
\bibitem{BZ1}
W. A. Bardeen and B. Zumino, Nucl. Phys. B244 (1984), 421.
\bibitem{Jack1}
R. Jackiw, in: ``Current Algebras and Anomalies'', eds. S. B. Treiman,
R. Jackiw, B. Zumino, and E. Witten, World Scientific, Singapore 1985.
\bibitem{Fad1}
L. D. Faddeev, Phys. Lett. B145 (1984), 81.
\bibitem{Mick1}
J. Mickelsson, Lett. Math. Phys. 7 (1983), 45; Comm. Math. Phys.
97 (1985), 361.
\bibitem{Sto1}
R. Stora, in: ``Recent Progress in Gauge Theories'', ed. H. Lehmann,
1983 Cargese Lectures, Nato ASI Series, Plenum Press, NY 1984.
\bibitem{Zu1}
B. Zumino, in: ``Relativity, Groups and Topology 2'', eds. B. S. DeWitt
and R. Stora, 1983 Les Houches Lectures, North Holland, Amsterdam 1984.
\bibitem{Tsu}
I. Tsutsui, Phys. Lett. B229 (1989), 51.
\bibitem{Keln1}
G. Kelnhofer, Z. Phys. C52 (1991), 89.
\bibitem{Pawl2}
J. Pawlowski, Phys. Rev. D 57 (1998), 1193.
\bibitem{ABH}
C. Adam, R. A. Bertlmann, and P. Hofer, Riv. Nuovo Cim. 16 (1993), 1.
\bibitem{COMANOM}
C. Adam, Ann. Phys. 265 (1998) 198.
\bibitem{Bertl1}
R. A. Bertlmann, ``Anomalies in Quantum Field Theory'', Clarendon Press,
Oxford 1996.
\bibitem{Ek1}
C. Ekstrand, hep-th/9903147.

\end{thebibliography}
\end{document}